\documentclass[letter,11pt]{article}
\pdfoutput=1
 \usepackage{jheppub}
 
 \usepackage{graphicx}
 \usepackage{hyperref}
 \usepackage[utf8]{inputenc}
 \usepackage{amssymb}
\usepackage{multirow}
\usepackage{soul}
 
 \usepackage{booktabs}
\usepackage{mathrsfs}
\usepackage{epsfig}
\usepackage{graphicx}
\usepackage{subfigure}
\usepackage{dcolumn}
\usepackage{bm}
\usepackage{amsmath}
\usepackage{slashed}
\usepackage{multirow}
\usepackage{color}
\usepackage{dcolumn}
\usepackage{bm}
\usepackage{color}

\allowdisplaybreaks

 \newcommand{\lsim}{{\;\raise0.3ex\hbox{$<$\kern-0.75em\raise-1.1ex\hbox{$\sim$}}\;}}
\newcommand{\gsim}{{\;\raise0.3ex\hbox{$>$\kern-0.75em\raise-1.1ex\hbox{$\sim$}}\;}}
\newcommand{\beq}{\begin{equation}}
\newcommand{\eeq}{\end{equation}}
\newcommand{\bea}{\begin{eqnarray}}
\newcommand{\eea}{\end{eqnarray}}

\def\baa{\begin{array}}
\def\eaa{\end{array}}

\mathchardef\minus="002D

\preprint{ }

\title{Vacuum stability of the type II seesaw leptogenesis from inflation}

\author{Chengcheng Han$^1$,}
\emailAdd{hanchch@mail.sysu.edu.cn}

\author{Sihui Huang$^{1}$, }
\emailAdd{huangsh99@mail2.sysu.edu.cn}

\author{Zhanhong Lei$^{1}$}
\emailAdd{leizhh3@mail2.sysu.edu.cn}

\vspace{0.5cm}

\affiliation{$^1$School of Physics, Sun Yat-Sen University, Guangzhou 510275, P. R. China}

\vspace{0.5cm}

\abstract{
Recently it has been found that introducing a triplet Higgs to the standard model could provide a feasible leptogenesis to generate the baryon asymmetry of our universe, providing that the inflation is driven by the mixing state of the triplet Higgs and SM Higgs. In this work, we survey the viable parameter space satisfying the vacuum stability and perturbativity in this model. We find that the introduction of the triplet Higgs would also ameliorate the problem of the Higgs vacuum instability.  We present two representative parameter regions where the origin of neutrino masses, baryon asymmetry of the universe as well as inflation can be explained while keeping consistent with the condition of vacuum stability and perturbativity.}

\makeatletter
\def\@fpheader{\relax}
\makeatother

\date{2020.06.15}

\begin{document} 
\maketitle
\flushbottom
\newpage

\section{Introduction}

The Standard Model(SM), a theory for strong and electroweak interactions, is regarded as a landmark of modern particle physics. However, there remain several unsolved problems in particle physics and cosmology, such as the origin of the cosmic inflation~\cite{Brout:1977ix, Sato:1980yn, Guth:1980zm, Linde:1981mu, Albrecht:1982wi}, neutrino masses, and baryon asymmetry of our universe. Numerous models have been promoted to solve one or two of the above problems but rarely solve all of them at once. Interestingly, recently the authors found that the type II seesaw model~\cite{Magg:1980ut, Cheng:1980qt, Lazarides:1980nt, Mohapatra:1980yp} could provide a simple framework to address all three problems simultaneously~\cite{Barrie:2021mwi, Barrie:2022cub}. The idea is that the neutrino masses are generated by the traditional type II seesaw mechanism where the neutrino masses are provided by the vacuum value of the neutral part of the triplet Higgs, and the inflation is driven by the mixing state of the triplet Higgs and SM Higgs doublet in the early universes, and the baryon asymmetry is generated by the Affleck-Dine mechanism~\cite{Affleck:1984fy, Dine:1995kz} during the inflation. 

However, since the model contains many additional scalars, it is important to examine whether such a model satisfies various theoretical constraints, particularly, the stability of the vacuum and the perturbativity of the parameters. Actually, the standard model Higgs itself already suffers from the instability problem due to the second minimum of the potential at the high energy scale~\cite{Buttazzo:2013uya, Han:2018yrk}. It would be interesting to study whether this problem can be addressed by the introduction of the triplet Higgs. Although there is already a lot of work concerning the vacuum stability and perturbativity in the framework of type II seesaw~\cite{Arhrib:2011uy, Chun:2012jw, BhupalDev:2013xol, Bonilla:2015eha, Haba:2016zbu, Moultaka:2020dmb}, many of them focus on these problems around the electroweak scale. However, in the model of type II seesaw triplet leptogenesis~\cite{Barrie:2021mwi, Barrie:2022cub}, the triplet Higgs also plays a role of inflaton. If a second vacuum develops at a higher energy scale, we could possibly either worry about the transition of our vacuum into the second vacuum or that we keep staying in the second vacuum from the beginning. The reason for the latter statement is that after the inflation, there exists an oscillation stage of the triplet Higgs and SM Higgs doublet~\cite{Barrie:2021mwi, Barrie:2022cub}, and then there is a high chance that the oscillation would end at the second vacuum. One may wonder whether the thermal effect during the radiation stage following reheating could save this situation. However, it is believed that the reheating temperature could not be much higher than $\sim 10^{14}$ GeV~\cite{Sfakianakis:2018lzf, Barrie:2021mwi}. If the second minimum develops at a higher scale, we would still keep staying in the second vacuum. In this paper, to simplify our analysis we require the electroweak vacuum to be absolutely stable and no second vacuum develops at any intermediate scale between the electroweak scale and Planck scale. 

In addition, to generate the baryon asymmetry, the inflation should be driven by a mixing state of the triplet Higgs and the SM Higgs doublet, providing additional conditions on the viable parameter space. Therefore it would be intriguing to give a full analysis of the parameter space satisfying the vacuum stability and perturbativity, keeping the above three problems being resolved.

The paper is organized as follows. In Sec.~\ref{leptogenesis} we overview the idea of the leptogenesis from Higgs inflation. 
In Sec.~\ref{vacuum} we summarize the theoretical requirement of vacuum stability as well as the necessary condition for Higgs inflation with successful leptogenesis. We show our numerical result in Sec.~\ref{numerical} and we draw the conclusion in Sec.~\ref{conclusion}.

\section{Leptogenesis from Higgs inflation}
\label{leptogenesis}
In this section, we briefly overview how leptogenesis could occur if the inflation is provided by a mixture of triplet Higgs and SM Higgs doublet. We will first show the condition of such mixture to drive the inflation and then present the Affleck-Dine mechanism to generate the baryon asymmetry during the inflation stage.   
\subsection{Inflation from Higgs}
The Type-II seesaw mechanism introduces a triplet scalar $\Delta$ to SM, which transforms as $(1,3,2)$ under the $SU(3)_c\times SU(2)_L\times U(1)_Y$ gauge group and carries a lepton number $l_\Delta=-2$. Under tensor representation of $SU(2)_L$, the triplet $\Delta$ and SM Higgs field can be written as:
\begin{eqnarray}
    \label{con:field}
    \Delta=\left(
    \begin{matrix}
        \Delta^+/\sqrt{2} & \Delta^{++}\\
        \Delta^0 & -\Delta^+/\sqrt{2}
    \end{matrix}
    \right)~~~
    H=\left(
    \begin{matrix}
        h^+ \\ h
    \end{matrix}
    \right)~.
\end{eqnarray}
The potential of the Higgs fields is given by:
\begin{eqnarray}
\label{con:potential}
    V(H,\Delta)&=&-m^2_HH^\dagger H+m_\Delta^2\text{Tr}(\Delta^\dagger\Delta) +\lambda_H(H^\dagger H)^2+\lambda_1(H^\dagger H)\text{Tr}(\Delta^\dagger\Delta) \nonumber \\
    &&+\lambda_2(\text{Tr}(\Delta^\dagger\Delta))^2 +\lambda_3\text{Tr}(\Delta^\dagger\Delta)^2+\lambda_4H^\dagger\Delta\Delta^\dagger H +[\mu(H^Ti\sigma^2\Delta^\dagger H) \nonumber \\
   && +\frac{\lambda_5}{M_p}(H^Ti\sigma^2\Delta^\dagger H)(H^\dagger H)+ \frac{\lambda_5^\prime}{M_p}(H^Ti\sigma^2\Delta^\dagger H)(\Delta^\dagger \Delta) +h.c.]~.
\end{eqnarray}
where the terms in the bracket break the $U(1)_L$ symmetry and we also include the Planck suppressed operators. Even if the high dimension operators have no effect on the low energy physics, they could dominate the $U(1)_L$ breaking terms during inflation. 
The Yukawa term becomes
\begin{eqnarray}
    \mathcal{L}_{Yukawa}=\mathcal{L}^{SM}_{Yukawa}-\frac{1}{2}Y_{\nu jk}\Bar{L}^c_ji\sigma^2\Delta L_k+h.c.~.
\end{eqnarray}
However, the recent CMB observation already exclude the possibility of the potential being a simple polynomial function. The reason is that the limit of  tensor to scalar ratio around 0.056 indicates that the potential of the inflaton should be flat enough at which inflation happens. Fortunately, it is known that in Higgs inflation a non-minimal coupling of Higgs field to gravity can flatten the potential and result in a Starobinsky type model. Therefore we extend the model with both H and $\Delta$ taking the non-minimal coupling to gravity, and then the Lagrangian is as follows:
\begin{eqnarray}
    \label{con:lagrangian}
    \begin{aligned}
    \frac{\mathcal{L}}{\sqrt{-g}}=&-\frac{1}{2}M^2_PR-\xi_HH^\dagger HR-\xi_\Delta \text{Tr}(\Delta^\dagger \Delta)R\\
    &-g^{\mu\nu}(D_\mu H)^\dagger(D_\nu H)-g^{\mu\nu}\text{Tr}(D_\mu \Delta)^\dagger(D_\nu \Delta)-V(H,\Delta)+\mathcal{L}_{Yukawa}~.
    \end{aligned}
\end{eqnarray}

To simplify our analysis, we only focus on the neutral components of $\Delta$ and $H$. The relevant lagrangian becomes:
\begin{eqnarray}
\begin{aligned}
    \frac{\mathcal{L}}{\sqrt{-g}}=&-\frac{1}{2}M^2_PR-\xi_H|h|^2R-\xi_\Delta |\Delta^0|^2R-(\partial_\mu h)^2-(\partial_\mu \Delta^0)^2\\
    &-V(h,\Delta^0)-\lbrack \frac{1}{2}Y_{\nu ij}\Bar{\nu}^c_i\nu_j\Delta^0+h.c.\rbrack+\cdots~, 
\end{aligned}
\end{eqnarray}
where $V(h,\Delta^0)$ is defined as
\begin{eqnarray}
    V(h,\Delta^0)&=&-m_H^2|h|^2+m_\Delta^2|\Delta^0|^2+\lambda_H|h|^4+\lambda_\Delta |\Delta^0|^4+\lambda_{H\Delta}|h|^2|\Delta^0|^2\nonumber \\
   &&  -\left(\mu h^2 {\Delta^0}^* + \frac{\lambda_5}{M_p} |h|^2  h^2  {\Delta^0}^*  + \frac{\lambda^\prime_5}{M_p} |\Delta^0|^2 h^2  {\Delta^0}^* +h.c. \right) +...  ~,
\end{eqnarray}
where $\lambda_\Delta=\lambda_2+\lambda_3,~\lambda_{H\Delta}=\lambda_1+\lambda_4$. The Yukawa term of lepton and $\Delta^0$ provides a Majorana mass term for neutrino once $\Delta$ obtains a non-vanishing vacuum expectation value.

We parameterize the $h$ and  $\Delta^0$ by the polar coordinates: $h\equiv\frac{1}{\sqrt{2}}\rho_He^{i\eta}$ and $\Delta^0\equiv\frac{1}{\sqrt{2}}\rho_\Delta e^{i\theta}$. Then the Lagrangian can be written as:
\begin{eqnarray}
    \label{Lagrang1}
    \frac{\mathcal{L}}{\sqrt{-g}}=-\frac{1}{2}(M_P^2+\xi_H\rho_H^2+\xi_\Delta\rho_\Delta^2)R-\frac{1}{2}(\partial_\mu \rho_H)^2-\frac{1}{2}(\partial_\mu \rho_\Delta)^2-V(h,\Delta^0)+\cdots~.
\end{eqnarray}
Considering $\xi_H\rho_H^2+\xi_\Delta\rho_\Delta^2\gg 1$ during inflation, we can perform a weyl transformation:
\begin{eqnarray}
    \label{con:transform}
    g_{E\mu\nu}=\Omega^2 g_{\mu\nu},~~\Omega^2=1+\frac{\xi_H\rho_H^2}{M_P^2}+\frac{\xi_\Delta\rho_\Delta^2}{M_P^2}~.
\end{eqnarray}
Then Lagrangian is translated from Jordan frame into Einstein frame:
\begin{eqnarray}
    \label{con:einstein}
    \frac{\mathcal{L}}{\sqrt{-g_E}}=-\frac{1}{2}M_P^2R_E-\frac{1}{2\Omega^2}\left((\partial_\mu \rho_H)^2+(\partial_\mu \rho_\Delta)^2\right)-3M_P^2(\partial_\mu \text{log}\Omega)^2-\frac{V(h,\Delta^0)}{\Omega^4}+\cdots~.
\end{eqnarray}
Subsequent derivations in this section mainly follows~\cite{Lebedev:2011aq}. Here we use the natural unit $M_p= 1$. Then from Eq.~\ref{con:transform} and Eq.~\ref{con:einstein} we get:
\begin{eqnarray}
    \label{con:kinetic}
    \mathcal{L}_{kin}=-\frac{3}{4}\left(\partial_\mu \text{log}(\xi_H\rho_H^2+\xi_\Delta\rho_\Delta^2)\right)^2-\frac{1}{2(\xi_H\rho_H^2+\xi_\Delta\rho_\Delta^2)}\left((\partial_\mu \rho_H)^2+(\partial_\mu \rho_\Delta)^2\right)~.
\end{eqnarray}
We can redefine the fields:
\begin{eqnarray}
    \label{con:redefine}
    \chi=\sqrt{\frac{3}{2}}\text{log}(\xi_H\rho_H^2+\xi_\Delta\rho_\Delta^2),~~\kappa=\frac{\rho_H}{\rho_\Delta}~.
\end{eqnarray}
Plugging Eq.~\ref{con:redefine} into Eq.~\ref{con:kinetic} and requiring a large non-minimal coupling $\xi_H, \xi_\Delta\gg 1$, we get:
\begin{eqnarray}
    \mathcal{L}_{kin}=-\frac{1}{2}(\partial_\mu\chi)^2-\frac{1}{2}\frac{\xi_H^2\kappa^2+\xi_\Delta^2}{(\xi_H\kappa^2+\xi_\Delta)^3}(\partial_\mu\kappa)^2~.
\end{eqnarray}
For large field value, the quartic terms dominate the potential, thus the $\mu$ term in $V(h,\Delta^0)$ can be ignored. Considering a large $\chi$, the potential of the scalar fields can be written as 
\begin{eqnarray}
    U(\rho_H,\rho_\Delta)=\frac{V}{\Omega^4}=\frac{\lambda_H\kappa^4+\lambda_{H\Delta}\kappa^2+\lambda_\Delta}{4(\xi_H\kappa^2+\xi_\Delta)^2}~,
\end{eqnarray}
with the following minima:
\begin{eqnarray}
    \label{con:minima}
    &&(1)~2\lambda_H\xi_\Delta-\lambda_{H\Delta}\xi_H>0,~2\lambda_\Delta\xi_H-\lambda_{H\Delta}\xi_\Delta>0,~\kappa=\sqrt{\frac{2\lambda_\Delta\xi_H-\lambda_{H\Delta}\xi_\Delta}{2\lambda_H\xi_\Delta-\lambda_{H\Delta}\xi_H}}~,\\
    &&(2)~2\lambda_H\xi_\Delta-\lambda_{H\Delta}\xi_H>0,~2\lambda_\Delta\xi_H-\lambda_{H\Delta}\xi_\Delta<0,~\kappa=0~,\\
    &&(3)~2\lambda_H\xi_\Delta-\lambda_{H\Delta}\xi_H<0,~2\lambda_\Delta\xi_H-\lambda_{H\Delta}\xi_\Delta>0,~\kappa=\infty~,\\
    &&(4)~2\lambda_H\xi_\Delta-\lambda_{H\Delta}\xi_H<0,~2\lambda_\Delta\xi_H-\lambda_{H\Delta}\xi_\Delta<0,~\kappa=0,\infty~.
\end{eqnarray}
In scenario 1, the inflation is dominated by $\Delta^0$ and $H$, which is just the case we are interested in. Thus we get our first constraint on the parameters to ensure the inflation is driven by a mixing state of the triplet Higgs and SM doublet Higgs, i.e,
\begin{eqnarray}
    x=\frac{\xi_H}{\xi_\Delta},~2\lambda_H-x\lambda_{H\Delta}>0,~2x\lambda_\Delta-\lambda_{H\Delta}>0~.
\end{eqnarray}
Note that a necessary condition for vacuum stability is that $\lambda_\Delta > 0, \lambda_H > 0$. However, $\lambda_{H\Delta}$ could be either positive or negative. Obviously, if $\lambda_{H\Delta} < 0$, the above condition can be automatically satisfied, while for the case of  $\lambda_{H\Delta} > 0$, a dedicated study should be preformed and we leave the details of discussion in Sec.~\ref{numerical}.

\subsection{Leptogensis from Higgs inflation}
we can simplify our analysis by defining the inflaton as $ \varphi $, through the following relations to the polar coordinate fields,
\begin{eqnarray}
&&  \rho_{H} = \varphi \sin \alpha,~\rho_{\Delta} = \varphi \cos \alpha~,  \nonumber \\
&& \xi \equiv \xi_H \sin^2 \alpha + \xi_\Delta \cos^2 \alpha~.
\end{eqnarray}

The Lagrangian in Eq. (\ref{Lagrang1}) is now given by,
\begin{eqnarray}
\frac{\mathcal L}{\sqrt{-g}} = -\frac{M_p^2}{2}  R -\frac{\xi}{2}   \varphi^2  R  - \frac{1}{2} g^{\mu\nu} \partial_\mu \varphi \partial_\nu \varphi  -\frac{1}{2} \varphi^2 \cos^2\alpha~ g^{\mu\nu} \partial_\mu \theta \partial_\nu \theta-V(\varphi, \theta) ~,
\label{lagrang1}
\end{eqnarray}
where 
\begin{eqnarray}
\hspace{-0.5cm}V(\varphi, \theta)  = \frac{1}{2} m^2 \varphi^2 + \frac{\lambda}{4} \varphi^4 + 2\varphi^3 \left(\tilde \mu   + \frac{\tilde \lambda_5}{M_p} \varphi^2\right) \cos\theta ~,
\end{eqnarray}
and
\begin{eqnarray}
 && m^2 = m^2_{\Delta} \cos^2 \alpha  -m_H^2 \sin^2 \alpha~,  \nonumber \\
 && \lambda = \lambda_H \sin^4 \alpha + \lambda_{H\Delta} \sin^2 \alpha \cos^2 \alpha + \lambda_{\Delta} \cos^4 \alpha~,  \nonumber \\
 && \tilde \mu  = -\frac{1}{2\sqrt{2}}\mu \sin^2\alpha  \cos \alpha~,  \nonumber \\
 && \tilde \lambda_5  = - \frac{1}{4\sqrt{2}} (\lambda_5 \sin^4\alpha  \cos \alpha + \lambda^\prime_5 \sin^2\alpha  \cos^3 \alpha)~.
\end{eqnarray}

To fit the current CMB observation, we need roughly $\frac{\xi}{\sqrt{\lambda}} \approx 5\times 10^4$. The Affleck-Dine mechanism will be realised in our scenario through the motion of the dynamical field $\theta$. The size of the generated lepton asymmetry will be determined by the size of the non-trivial motion induced in $\theta$ sourced by inflation. During inflation, $m \ll \varphi$, which means that the quartic term in the Jordan frame potential dominates the inflationary dynamics. 
The lepton number asymmetry generated by the motion of the triplet Higgs phase $\theta$ at the end of inflation,
\begin{eqnarray}
{n_L}_\textrm{end}&=& Q_L \varphi^2_\textrm{end} \dot\theta_\textrm{end} \cos^2 \alpha  \nonumber \\
&\simeq& - Q_L \tilde \lambda_5 \varphi^3_\textrm{end} \sin \theta_{\rm end} /\sqrt{3 \lambda}   ~.
\label{nlend1}
\end{eqnarray}
After inflation, the lepton number density is roughly red-shifted by $a^3$. To generate the correct baryon asymmetry of our univsere, we need ${n_L}_\textrm{end} \sim 10^{-16}$. 
For $\lambda\sim \mathcal O(0.1)$, $\varphi_\textrm{end}\sim 1$ in Planck unit, we just need  $\tilde \lambda_5 \sim 10^{-15}$.

\section{Vacuum stability from type II seesaw}
\label{vacuum}
In spite of the intriguing motivation, this mechanism suffers from constraints of theoretical consistency including vacuum stability, perturbativity, and the condition of inflation. The vacuum stability condition requires that the potential of Eq.~\ref{con:potential} does not develop a second vacuum between TeV scale and the Planck scale, otherwise our electroweak vacuum would be meta-stable and may transit to another vacuum. In the large field limit, we only need focus on the quartic term of the potential. These conditions can be summarized as Eq.~\ref{con:stability}, with detailed derivation in appendix \ref{derivation}.
\begin{eqnarray}
    \label{con:stability}
    \begin{aligned}
    C_1,C_2,C_3,C_4,C_5>0~\text{and}~[C_6>0~\text{or}~C^\prime_6>0]~.
    \end{aligned}
\end{eqnarray}
where
\begin{eqnarray}
    C_1&&=\lambda_H ~,\\
    C_2&&=\lambda_2+\lambda_3 ~,\\
    C_3&&=\lambda_2+\frac{1}{2}\lambda_3 ~,\\
    C_4&&=\lambda_1+2\sqrt{\lambda_H(\lambda_2+\lambda_3)} ~,\\
    C_5&&=\lambda_1+\lambda_4+2\sqrt{\lambda_H(\lambda_2+\lambda_3)} ~,\\
    C_6&&=|\lambda_4|\sqrt{\lambda_2+\lambda_3}-2\lambda_3\sqrt{\lambda_H} ~,\\
    C_6^\prime&&=2\lambda_1+\lambda_4+\sqrt{(8\lambda_H\lambda_3-\lambda_4^2)(2{\lambda_2}/{\lambda_3}+1)} ~.
\end{eqnarray}
Note that the above conditions are should be satisfied at any energy scale between TeV  and $M_p$. The perturbativity condition requires that all couplings should be not much beyond $O(1)$. Here we require all the couplings smaller than $\sqrt{4\pi}$.

Since inflation in our context requires a flat direction along the mixing of $h$ and $\Delta^0$, the condition in scenario 1 of Eq.~\ref{con:minima} should also be satisfied, which can be rewritten as:
\begin{eqnarray}
C_7>0,~ C_8>0~,
\end{eqnarray}
where
\begin{eqnarray}
~C_7&=&2\lambda_H-x\lambda_{H\Delta}~,  \\
~C_8&=&2x\lambda_\Delta-\lambda_{H\Delta}~, \\ 
x&=&\frac{\xi_H}{\xi_\Delta}~.
\end{eqnarray}
We stress that the conditions of $C_7>0$ and $C_8>0$ should be satisfied only at around Planck scale.

In our calculation we utilize SARAH \cite{Staub:2013tta} to derive the 2-loop renormalization group equations for the relevant couplings. The 1-loop result is extracted in appendix \ref{rge_oneloop} and we find it consistent with the result from~\cite{Bonilla:2015eha}. 

\section{Numerical results}
\label{numerical}

In our calculation, the free parameters include three gauge coupling parameters $g_1,~g_2,~g_3$, five quartic coupling parameters $\lambda_H,~\lambda_1,~\lambda_2,~\lambda_3,~\lambda_4$ and yukawa coupling of top quark and leptons with the triplet Higgs. The threshold effect of $\mu$ parameter can be ignored because it should be much smaller than GeV to avoid the wash out of the lepton asymmetry~\cite{Barrie:2021mwi, Barrie:2022cub}.  Here we set the triplet Higgs mass to be 1 TeV, then all the SM parameters firstly have to run from the top mass scale to TeV scale, then all the couplings run into Planck scale. During the intermediate scale, the vacuum stability condition has been checked and the inflation conditions $C_7$ and $C_8$ are only imposed at Planck scale. Since all the couplings, particularly the Higgs self-coupling is sensitive to the top mass, here we set the top mass to be 172 GeV and Higgs mass to be 125 GeV. All the SM model couplings at the top mass scale are derived following the formulas from~\cite{Buttazzo:2013uya}.

Before going to the numerical results, we give some analysis on the preference of parameter space. First of all, to realize the mixing of the triplet Higgs and SM Higgs doublet, we have the following relation,
\begin{equation}
    x=\frac{\xi_H}{\xi_\Delta},~2\lambda_H-x\lambda_{H\Delta}>0,~2x\lambda_\Delta-\lambda_{H\Delta}>0~.
\end{equation}
Since the vacuum stability requires $\lambda_H >0, \lambda_\Delta > 0$, if $\lambda_{H\Delta}< 0$ the above relation can be satisfied automatically. If $\lambda_{H\Delta} > 0$, we need both $2\lambda_H-x\lambda_{H\Delta}>0$ and $~2x\lambda_\Delta-\lambda_{H\Delta}>0$, imposing additional requirement for the parameter space.

In addition, it is known that the SM Higgs develops a second vacuum at the scale around $10^{10-11}$ GeV. However, due to the introduction of the new scalars, it is possible to remove the second vacuum by changing RGE running of SM Higgs self-coupling. From the RGE of $\lambda_H$
\begin{eqnarray}
(4\pi)^2\frac{d\lambda_H}{d\log \mu}&=&\frac{27}{200}g_1^4+\frac{9}{20}g_1^2g_2^2+\frac{9}{8}g_2^4+3\lambda_1^2+3\lambda_1\lambda_4+\frac{5}{4}\lambda_4^2-\frac{9}{5}g_1^2\lambda_H-9g_2^2\lambda_H  \nonumber \\
&&+24\lambda_H^2+12\lambda_Hy_t^2-6y_t^4,
\end{eqnarray}
we find that there are additional contributions from the term of $3\lambda_1^2+3\lambda_1\lambda_4+\frac{5}{4}\lambda_4^2$. Note that this term is always positive and it is possible to lift the $\lambda_H$ away from negative at a high energy scale if $\lambda_1,\lambda_4$ are large enough. We find that there are two typical scenarios, one is that the couplings of  $\lambda_1$, $\lambda_4$ are small and  $\lambda_H$ can be lift to be positive but keeps small at Planck scale($\mathcal O(0.01)$), and the other case is that $\lambda_1$, $\lambda_4$ are relatively large and final $\lambda_H$ is significantly lifted and becomes rather large at Planck scale($\mathcal O(1)$). In the following we will present the parameter space for these two typical scenarios. 

\begin{figure}[htbp]
    \centering
    \subfigure[$x=1$]{\includegraphics[height=0.27\textheight]{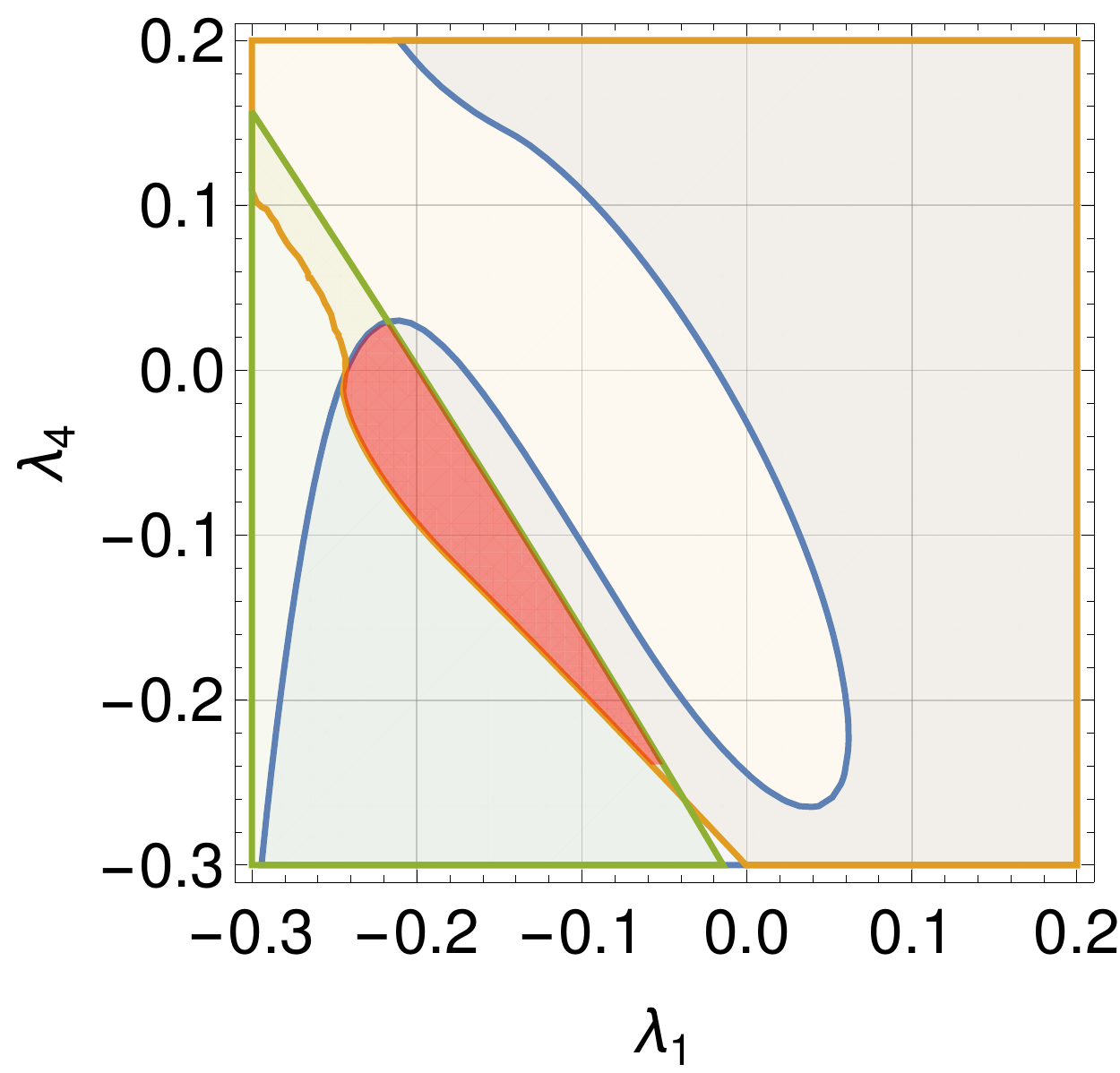}}
    \subfigure[$x=1/3$]{\includegraphics[height=0.27\textheight]{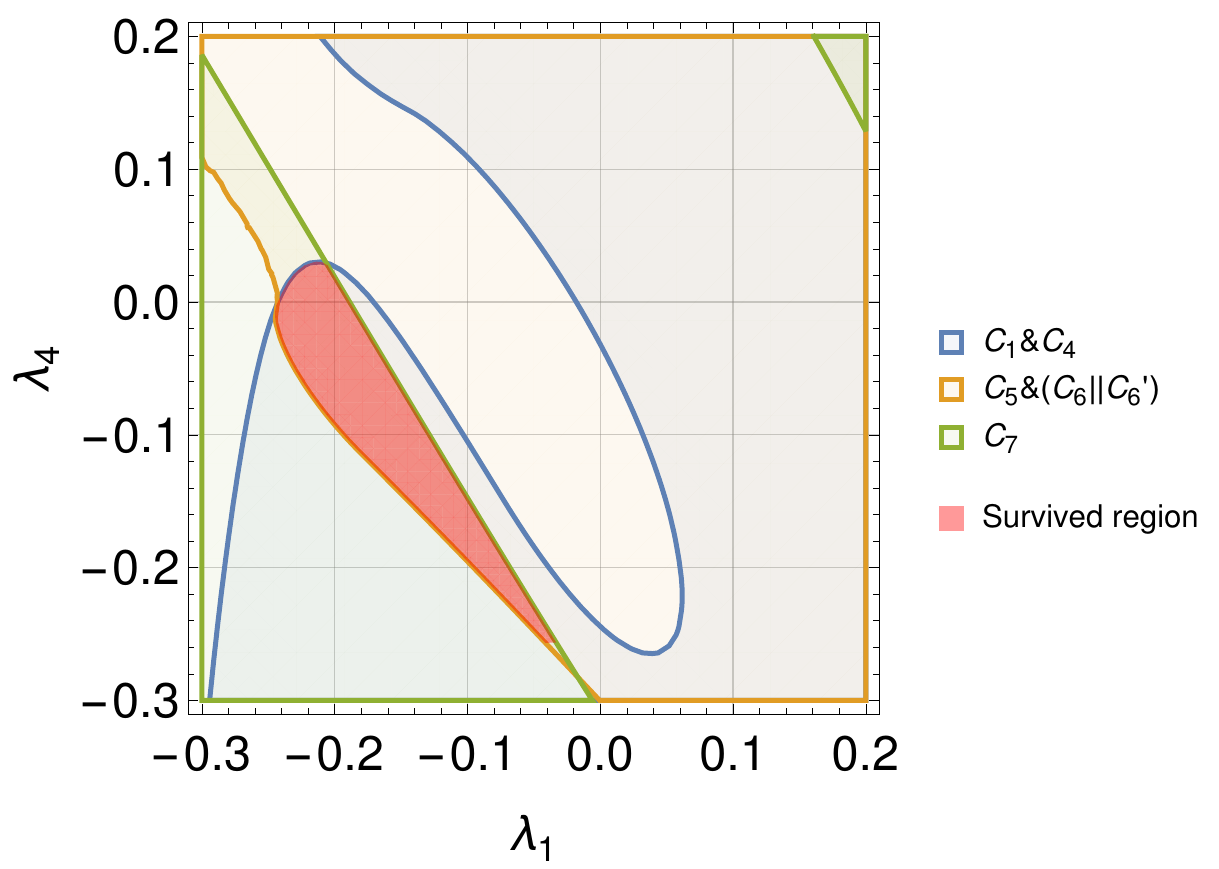}}
    \caption{\label{fig:case1}Regions surviving the conditions in the parameter space for $\lambda_2=0.15,\lambda_3=0.1,y_\nu=0$ at TeV scale.}
\end{figure}

In Fig.~\ref{fig:case1} we show the parameter region of a small $\lambda_H$ predicted at $M_p$ scale. Here we fixed $\lambda_2=0.15, \lambda_3=0.1$ at 1 TeV scale. For the yukawa coupling from the triplet Higgs and leptons, we assume that it is small enough and has no effect on the RGE running of other parameters and we will discuss its effect later. From left to right, $x$ is fixed to be $1, 1/3$ respectively. Note that the the condition $C_2>0$ and $C_3>0$ can be easily satisfied by set up. The constraints from other conditions are also shown in combination. Each of the combined conditions divide the parameter region into two parts and only one of them is allowed, as indicated by the allowed region(red). Different choices of $x$ has small effect and it affects slightly condition $C_7>0$. The reason is that $\lambda_H$ is small at high energy scale and at the boundary of the $C_7>0$ the $2\lambda_H-x\lambda_{H\Delta}$ approaches 0, therefore varying the $x$ value will slightly change  the boundary due to the smallness of $\lambda_H$. We note here that if the initial value of $\lambda_2$ becomes smaller, the parameter space would fast vanish. The reason that is that if we change $\lambda_2$ into a smaller value, both $\lambda_1$ and $\lambda_4$ would becomes larger and the condition $C_7>0$ would be easily violated.
\begin{figure}[htbp]
    \centering
    \includegraphics[width=0.7\linewidth]{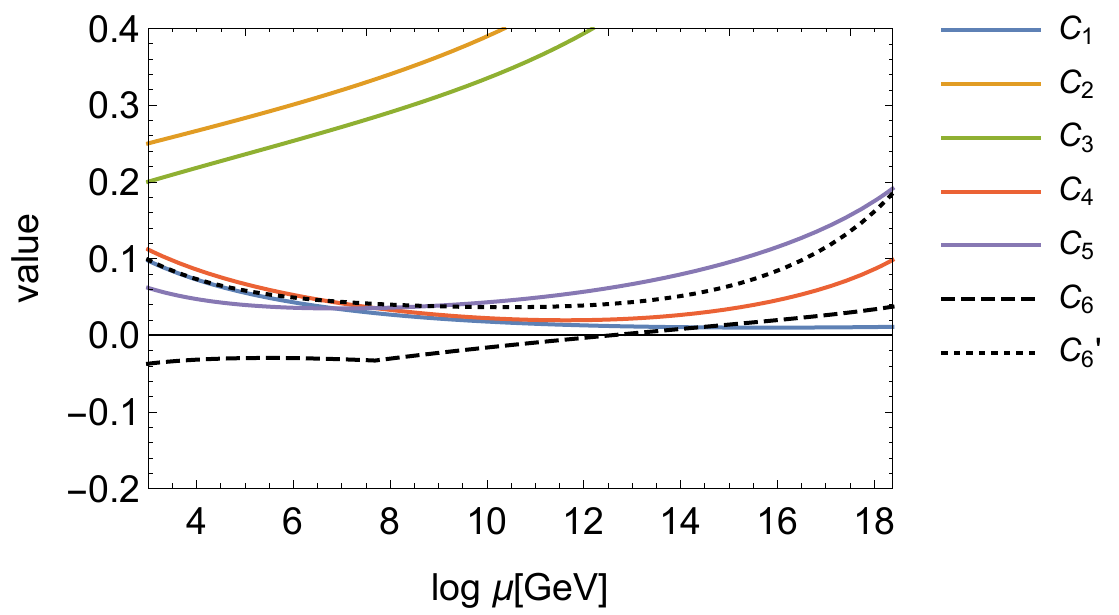}
    \caption{\label{fig:RGEcase1} Running of couplings from TeV scale to $M_{p}$, with $\lambda_1=-0.2,~\lambda_2=0.15,~\lambda_3=0.1,~\lambda_4=-0.05$ at TeV scale.}
\end{figure}

Here we give one benchmark in this parameter space. At TeV scale, the input parameters are fixed as follows: $\lambda_1=-0.2,~\lambda_2=0.15,~\lambda_3=0.1,~\lambda_4=-0.05$. At $M_{p}$ scale the parameters become  $\lambda_1=-0.14,~\lambda_2=0.97,~\lambda_3=0.33,~\lambda_4=0.093$.  Now we find $\lambda_H=0.011, ~\lambda_\Delta=1.3, ~\lambda_{H\Delta}=-0.046$. For the case $x=1$, the mixing angle of the Higgs and triplet Higgs during inflation and the effective $\lambda$ can be found to be
\begin{eqnarray}
\tan=\alpha =6.27,~~
\lambda=9.95\times 10^{-3}~.
\end{eqnarray}
To check the consistence of the vacuum stability, in Fig.~\ref{fig:RGEcase1} we show the RGE running of the vacuum stability conditions for this benchmark point. It is easy to find all the conditions are satisfied. Even if $C_6^\prime$ is negative at certain scale, $C_6$ is always keeping positive.

\begin{figure}[htbp]
    \centering
    \subfigure[$y_\nu=0,x=1$]{\includegraphics[height=0.27\textheight]{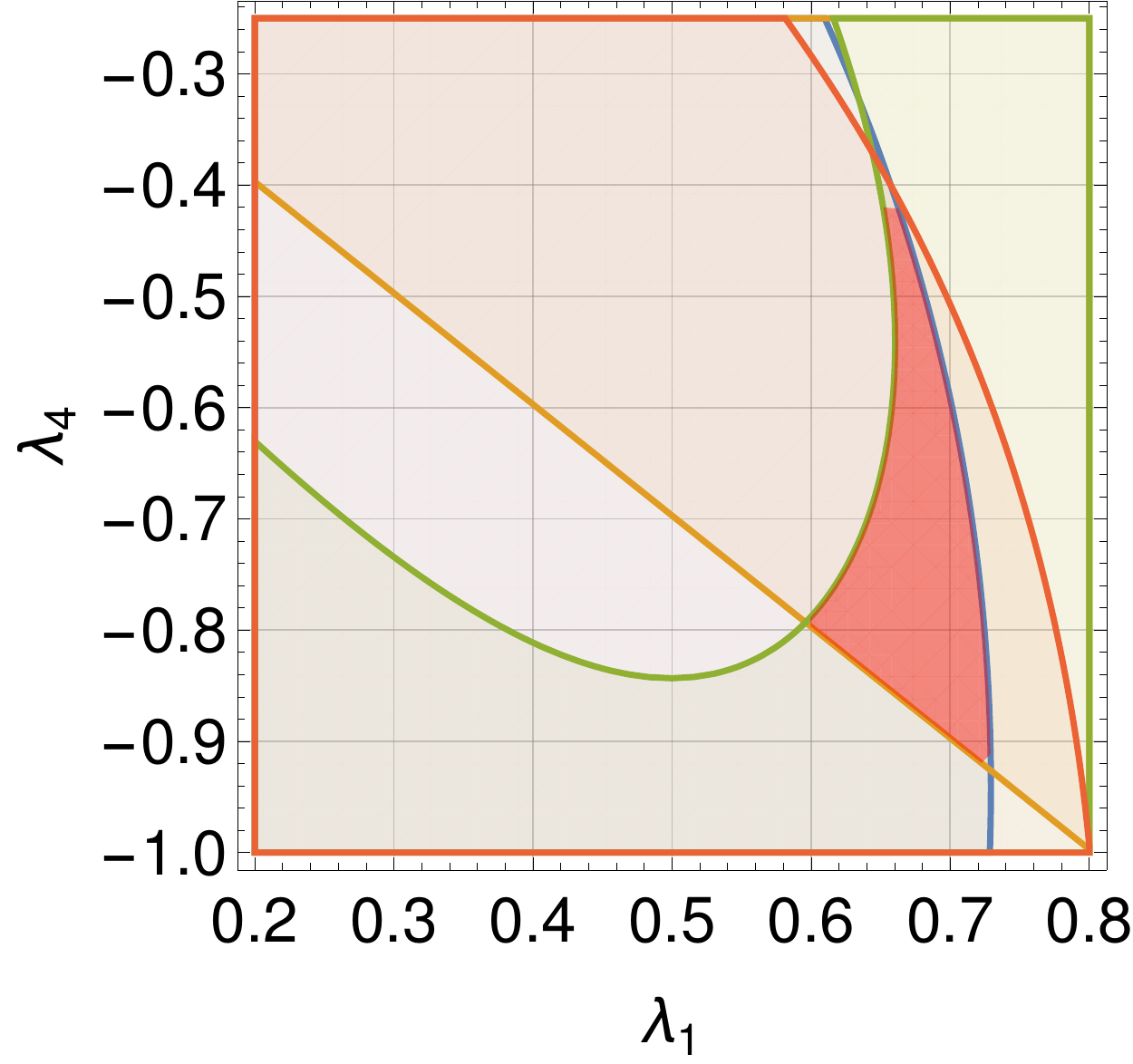}}
    \subfigure[$y_\nu=0,x=1/3$]{\includegraphics[height=0.27\textheight]{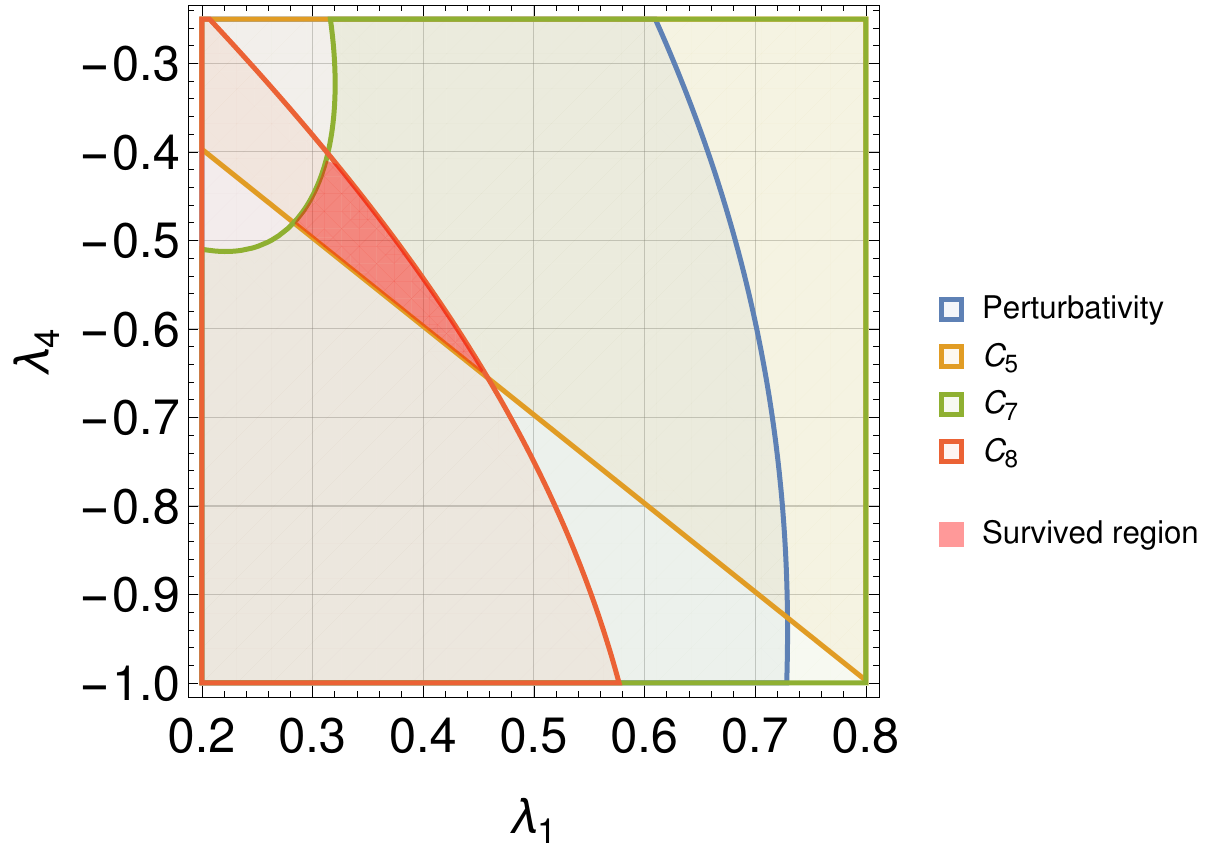}}
    \caption{\label{fig:case2}  Regions surviving the conditions in the parameter space for $\lambda_2=0,\lambda_3=0.1,y_\nu=0,x=1,~\frac{1}{3}$ at TeV scale.}
\end{figure}

In Fig.~\ref{fig:case2} we show the parameter region with $\lambda_{H}$ is $\mathcal{O}(1)$ at $M_p$ scale. Now we fixed $\lambda_2=0,\lambda_3=0.1$ at TeV scale. Again here we assume that the yukawa coupling is small enough and $x$ is fixed to be $1, 1/3$ respectively. It shows that the survived region is mainly determined by conditions $C_5>0$, $C_7>0$ and $C_8>0$. 
We also impose a perturbativity condition that all the parameters should be smaller than $\sqrt{4\pi}$.
Given the relatively large initial values of $\lambda_1~\text{and}~|\lambda_4|$ in Fig.~\ref{fig:case2}~(a), the perturbativity condition also matters. For a smaller $x$ as in Fig.~\ref{fig:case2}~(b), condition $C_7>0$ is relaxed but condition $C_8>0$ is strengthened, but the boundary of $C_5>0$ remains unchanged while the boundaries of $C_7>0$ and $C_8>0$ shift to left. For the case of $x=3$, we did not identify any parameter region since the condition $C_7>0$ becomes too strong.

\begin{figure}[htbp]
    \centering
    \includegraphics[width=0.7\linewidth]{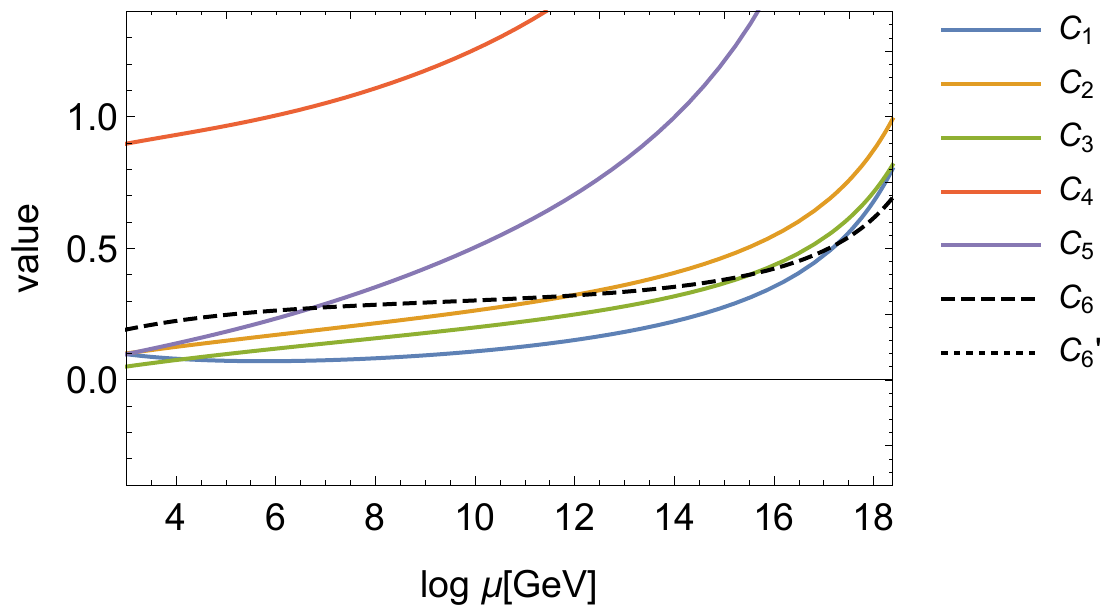}
    \caption{\label{fig:RGEcase2} Running of couplings from TeV scale to $M_{pl}$ scale, with $\lambda_1=0.7,~\lambda_2=0,~\lambda_3=0.1,~\lambda_4=-0.8$.}
\end{figure}

Again in Fig.~\ref{fig:RGEcase2} we show one benchmark point in the region of Fig.~\ref{fig:case2} for the RGE running of the vacuum stability conditions. The related parameters are fixed at $\lambda_1=0.7,~\lambda_2=0,~\lambda_3=0.1,~\lambda_4=-0.8$ at TeV scale. We find that at $M_{p}$ scale $\lambda_1=2.70,~\lambda_2=0.64,~\lambda_3=0.35,~\lambda_4=-1.32$ and  $\lambda_H=0.8, ~\lambda_\Delta=0.99, ~\lambda_{H\Delta}=1.38$. For $x=1$, we have $2\lambda_H-\lambda_{H\Delta}=0.22, ~2\lambda_\Delta-\lambda_{H\Delta}=0.60$. Then we can calculate the mixing angle and effective quartic coupling during inflation is  
\begin{equation}
\notag
\text{tan}~\alpha =1.65,~~
\lambda=0.77~.
\end{equation}

\begin{figure}[htbp]
    \centering
    \includegraphics[width=0.7\linewidth]{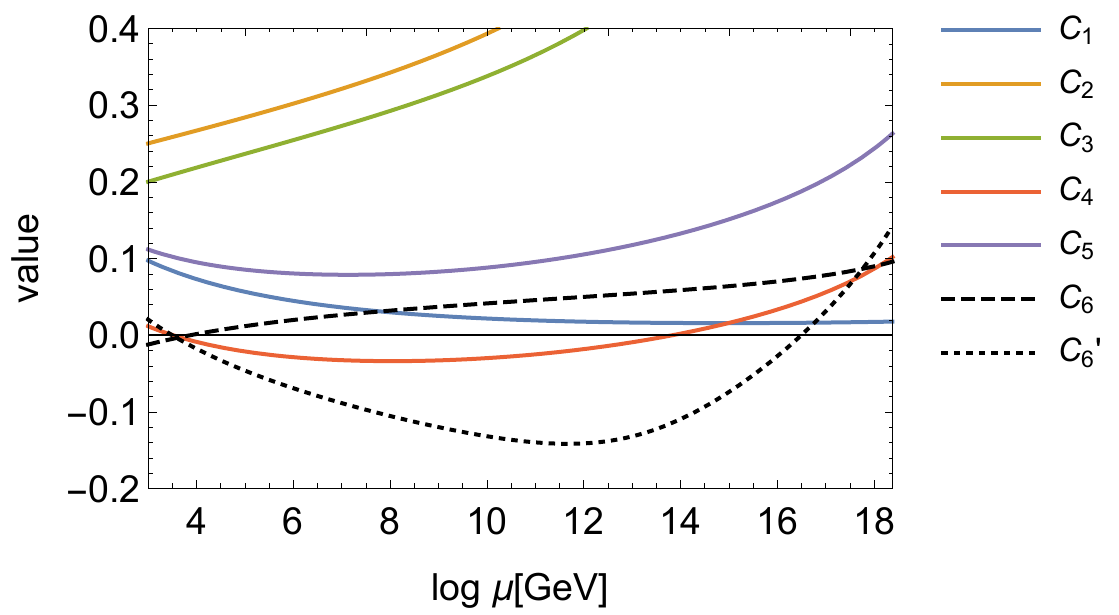}
    \caption{\label{fig:RGEcase3} Running of couplings from Electroweak scale to $M_{pl}$ scale, with $\lambda_1=-0.3,~\lambda_2=0.15,~\lambda_3=0.1,~\lambda_4=0.1$.}
\end{figure}

We stress the importance of imposing the vacuum stability conditions between low scale and $M_{p}$ since it is possible that the vacuum instability happens at an intermediate scale.
To illustrate this point, in Fig. \ref{fig:RGEcase3} we show a particular case where the vacuum stability conditions are satisfied at low scale and at Planck scale, but a second minimum develops at the scale around $10^{13}$ GeV. The couplings are fixed to  be $\lambda_1=-0.2,~\lambda_2=0.1,~\lambda_3=0.1,~\lambda_4=-0.06$ at TeV scale, and it clearly shows that vacuum stability conditions are violated between the scale of $10^4-10^{13}$ GeV.

\subsection{Effect from yukawa couplings of triplet Higgs with leptons}

In this subsection we discuss the effect of yukawa couplings from triplet Higgs with leptons. Since we have 6 independent yukawa couplings, to simplifiy our analysis, however, we only require one of them could be sizable and all the others are too small to have significant effect. 
In Fig.~\ref{fig:yukawa1} we show the effect of yukawa couplings. 

As shown in Appendix.~\ref{rge_oneloop}, the yukawa coupling $y_\nu$ contributes to the RGE of $\lambda_{H\Delta}$ from the term $\lambda_{H\Delta}y_\nu^2$. In the case of $\lambda_2=0.1,\lambda_3=0.1$, since $\lambda_{H\Delta}<0$ at almost all scales below $M_{pl}$, a non-vanishing yukawa coupling reduces the value of $\lambda_{H\Delta}$ at $M_{p}$, and then condition 7 can be relaxed slightly as shown in the left panel of Fig.~\ref{fig:yukawa1}. In the case of $\lambda_2=0.1,\lambda_3=0.1$, the $\lambda_1$ evolves with a positive and relatively large value. With the $\lambda_1y_\nu^2$ term in the RGE of $\lambda_1$, the value of $\lambda_1$ becomes larger at higher energy scale and it is easily beyond the perturbativity at $M_p$. Thus the perturbativity condition is strengthened with non-zero $y_\nu$. We can see that the survived region shrinks in as shown in the left panel of Fig.~\ref{fig:yukawa1} comparing with the plot without the contribution of yukawa coupling.

Notice that when $y_\nu \gtrsim 0.7$, it will hit the Landau pole before $M_p$ and such case should be avoided.

\begin{figure}[htbp]
    \centering
    \subfigure[$\lambda_2=0.1,\lambda_3=0.1$]{\includegraphics[height=0.25\textheight]{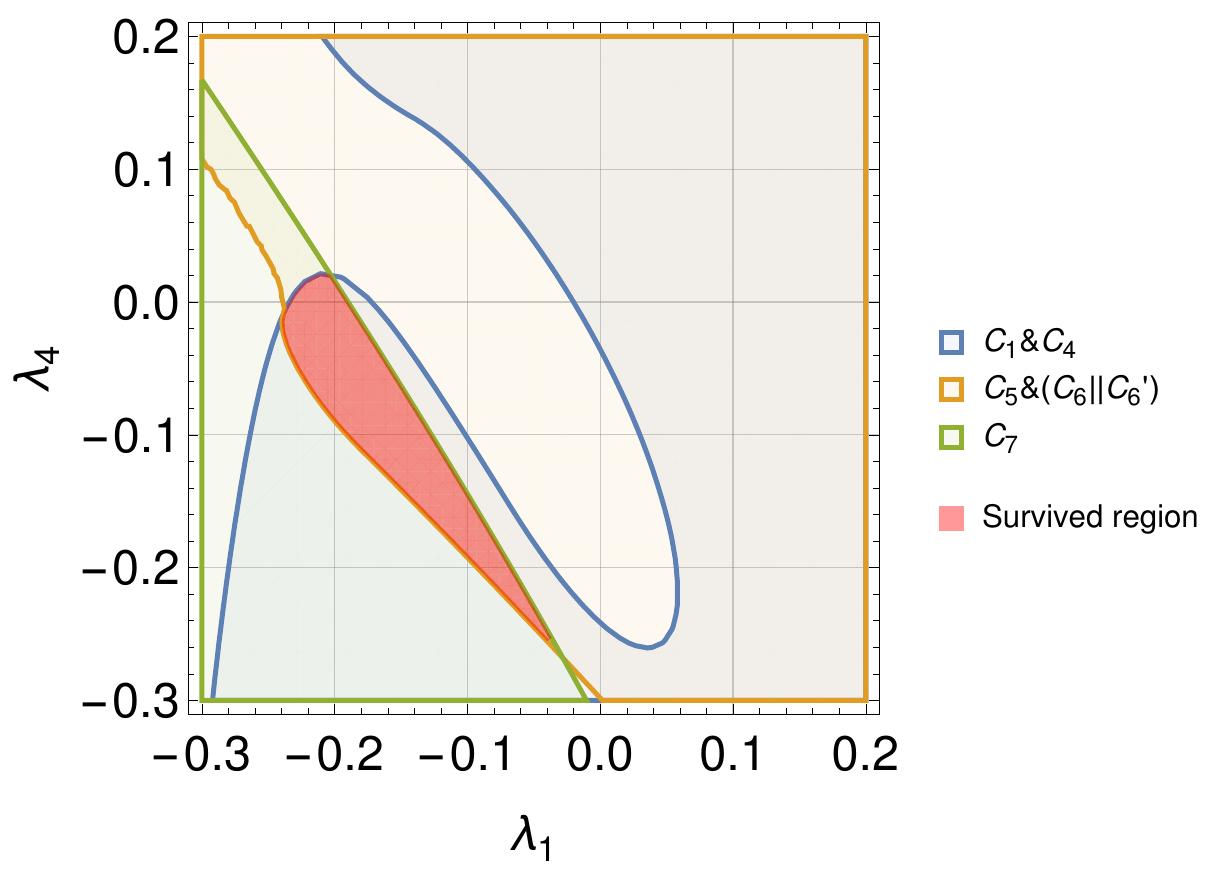}}
    \subfigure[$\lambda_2=0,\lambda_3=0.1$]{\includegraphics[height=0.25\textheight]{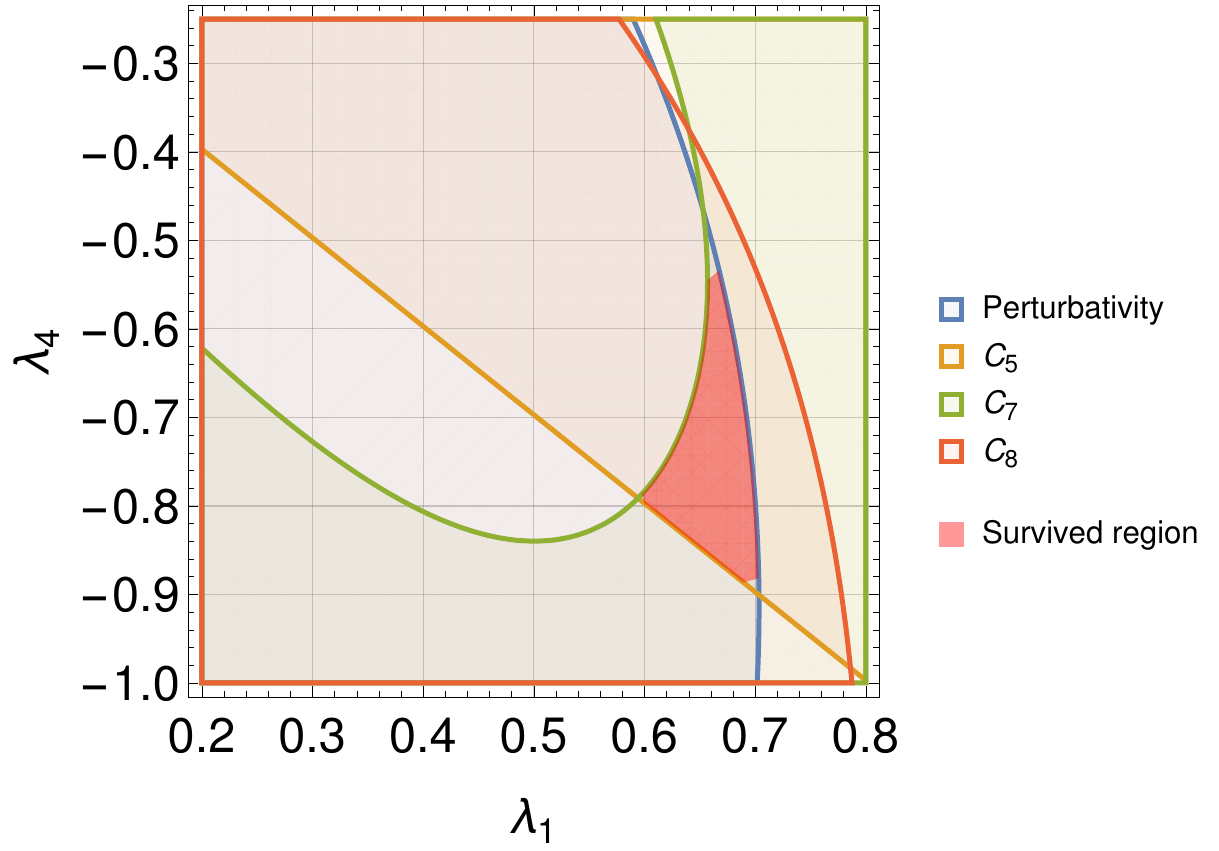}}    
    \caption{\label{fig:yukawa1} Regions surviving the conditions in the parameter space for $y_\nu=~0.6,x=1$.}
\end{figure}


\section{Phenomenological implications}

In the previous section we identify the parameter regions consistent with theoretical conditions. In this section we will show the low energy implications of the parameter space.  After electroweak symmetry spontaneous breaking, the triplet and SM Higgs field both obtain a vacuum expectation value,
\begin{equation}
    \label{eq:vev}
    \Delta=\left(
    \begin{matrix}
        0 & 0\\
        v_t/\sqrt{2} & 0
    \end{matrix}
    \right)~,~~
    H=\left(
    \begin{matrix}
        0 \\ v_d/\sqrt{2}
    \end{matrix}
    \right)~.
\end{equation}
Then the type II seesaw model predicts 7 massive particles: $H^{\pm\pm},~H^\pm,~H^0,~A^0,~\text{and}~h$.

\begin{figure}
    \centering
    \subfigure[$\lambda_2=0.15,\lambda_3=0.1,y_\nu=0$]{\includegraphics[width=0.48\linewidth]{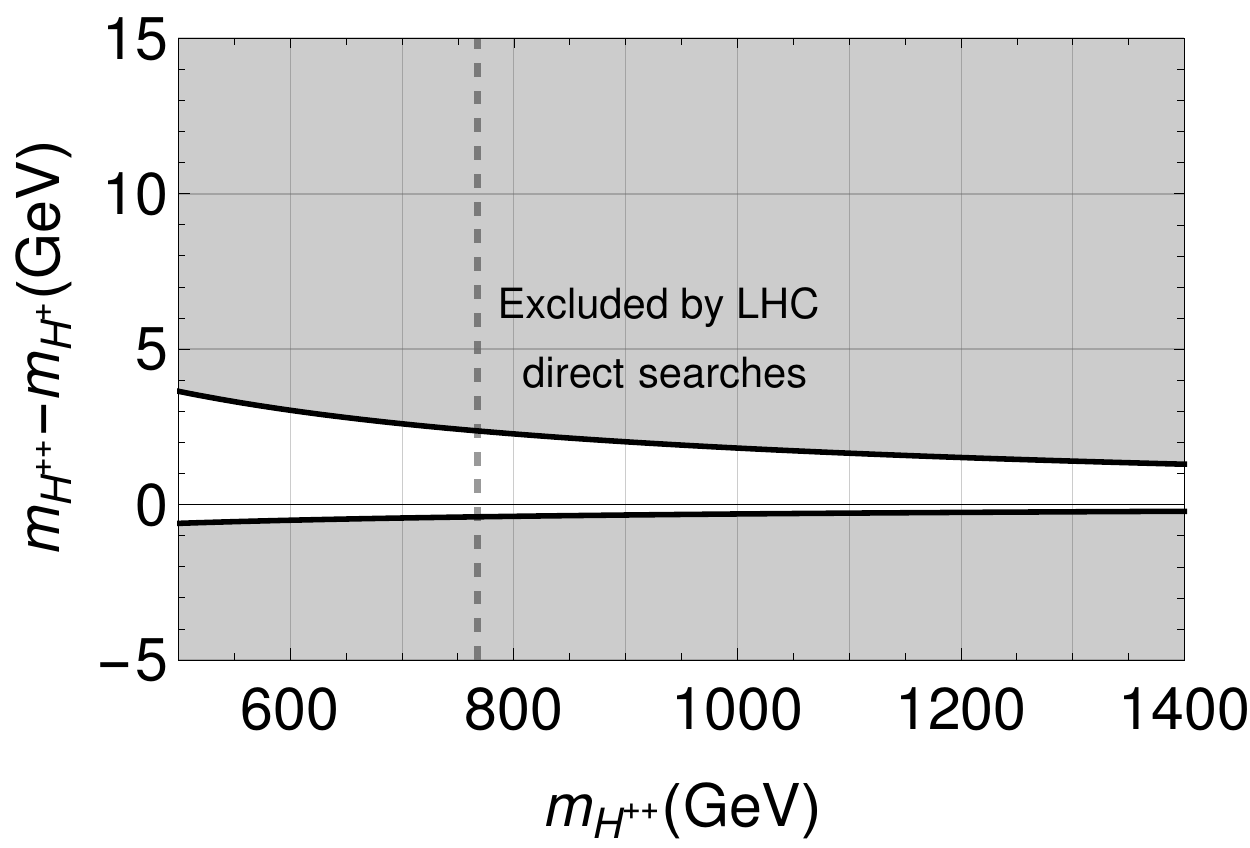}}
    \subfigure[$\lambda_2=0,\lambda_3=0.1,y_\nu=0$]{\includegraphics[width=0.48\linewidth]{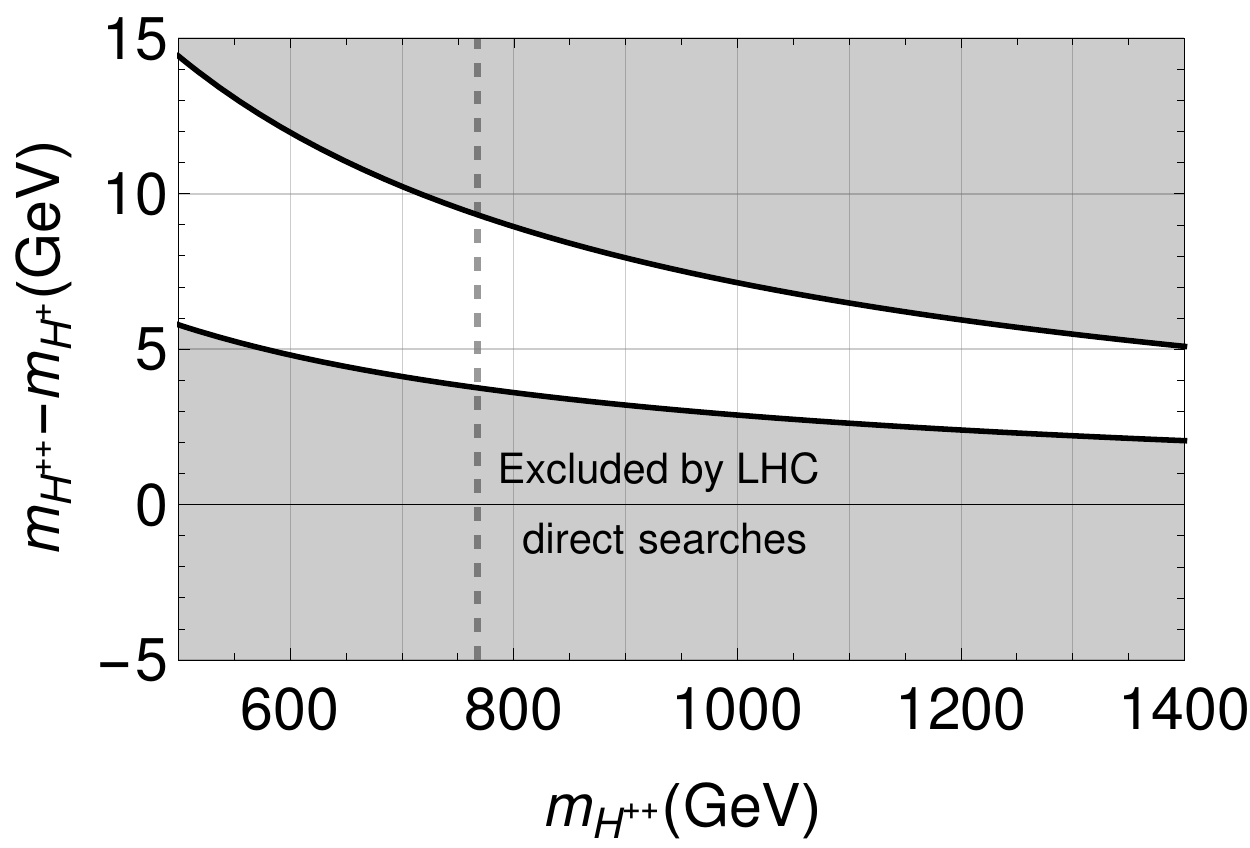}}
    \caption{\label{fig:mass_splitting} The mass splitting of $H^{\pm\pm}$ and $H^\pm$ predicted in the case of Fig.~\ref{fig:case1}  and Fig.~\ref{fig:case2} for $x=1$.}
\end{figure}

As in Ref.~\cite{Ashanujjaman:2021txz}, for $v_t\ll v_d$, which is the region we are interested in, the mass squared difference of these scalar particles can be deduced,
\begin{equation}
    \label{eq:mass_splitting}
    m^2_{H^{\pm\pm}}-m^2_{H^{\pm}}\simeq m^2_{H^{\pm\pm}}-m^2_{H^0/A^0}\simeq \frac{\lambda_4}{4}v_d^2~.
\end{equation}
Thus, $\lambda_4$ can be used to constrain the mass splitting of $H^{\pm\pm}$ and $H^\pm$, as shown in Fig.~\ref{fig:mass_splitting} (a) and Fig.~\ref{fig:mass_splitting} (b), which are related to the parameter regions in Fig.~\ref{fig:case1} and Fig.~\ref{fig:case2}, respectively.

We see that for a mass of doubly-charged Higgs around TeV, the mass splitting of doubly-charged Higgs and charged Higgs is generally below 10 GeV, thus the cascade decay of the doubly charged Higgs is highly suppressed. In addition, the vacuum value of the triplet Higgs is less than 10 keV, thus most of the doubly charged Higgs would decay into dileptons, providing a unique signature to test the model.

\section{Conclusions}
\label{conclusion}
In this work we investigate the parameter space of type II seesaw model consistent with the vacuum stability and perturbativity, while solving the problems of neutrino masses, baryon asymmetry of the universe and inflation. A successful leptogenesis  requires that the inflation is driven by the mixing state of the SM Higgs doublet and the triplet Higgs, providing additional constraints on the parameter space. After combining these conditions, we give two typical viable parameter regions where the SM Higgs self-coupling could be large $\mathcal{O}(1)$ or small $\mathcal{O}(0.01)$. 

\section{Acknowledgements}
We acknowledge Binglong Zhang for his helpful discussions. C. H. is supported by the Guangzhou Basic and Applied Basic Research Foundation under Grant No. 202102020885, and the Sun Yat-Sen University Science Foundation. 

\bibliographystyle{JHEP}
\bibliography{reference}

\appendix
\newpage
\renewcommand{\appendixname}{Appendix~\Alph{section}}

    \section{Derivation of the Vacuum Stability Conditions}
    \label{derivation}
   In this section, we calculate the vacuum stability conditions for a potential including the SM Higgs doublet and triplet Higgs. All the results can be found in \cite{Bonilla:2015eha, Moultaka:2020dmb} and here we just add some details. Here we consider the potential from Eq.~\ref{con:potential} where high dimension operators are neglected,
    \begin{equation}
    \label{potential}
    \begin{aligned}
     V(H,\Delta)=&-m^2_HH^\dagger H+m_\Delta^2\text{Tr}(\Delta^\dagger\Delta)+[\mu(H^Ti\sigma^2\Delta^\dagger H)+h.c.]\\
    &+\lambda_H(H^\dagger H)^2+\lambda_1(H^\dagger H)\text{Tr}(\Delta^\dagger\Delta)+\lambda_2(\text{Tr}(\Delta^\dagger\Delta))^2\\
    &+\lambda_3\text{Tr}(\Delta^\dagger\Delta)^2+\lambda_4H^\dagger\Delta\Delta^\dagger H+\cdots~.
    \end{aligned}
    \end{equation}
    It can be seen that in the large field limit the potential $V$ mainly depends on the following four items:$H^\dagger H$, $(\text{Tr}(\Delta^\dagger\Delta))^2$, $\text{Tr}(\Delta^\dagger\Delta)$ and  $H^\dagger\Delta\Delta^\dagger H$. Take $\text{Tr}(\Delta^\dagger\Delta)\ne0$ and define $r$, $\zeta$ and $\xi$ as the following non-negative dimensionless quantities,
    \begin{equation}
         H^\dagger H\equiv r\text{Tr}(\Delta^\dagger\Delta)~,
    \end{equation}
    \begin{equation}
        \text{Tr}(\Delta^\dagger\Delta)^2\equiv\zeta[\text{Tr}(\Delta^\dagger\Delta)]^2~,
    \end{equation}
    \begin{equation}
        H^\dagger\Delta\Delta^\dagger H\equiv\xi\text{Tr}(\Delta^\dagger\Delta)(H^\dagger H)~.
    \end{equation}
    For all scalar masses are positive, the potential does not classically drop to another minimum, but this still leaves the possibility of tunneling to a deeper minimum. In order for this not to happen, we must ensure that $V$ is bounded, which is equivalent to requiring that the quartic part of the potential in Eq.~\ref{potential}, $V^{(4)}$, must be positive for all non-zero field values.
    \begin{equation}
    \label{con:frac}
        \frac{V^{(4)}}{[\text{Tr}(\Delta^\dagger\Delta)]^2}=\lambda_Hr^2+\lambda_1r+\lambda_2+\lambda_3\zeta+\lambda_4\xi r~.
    \end{equation}
    For $V^{(4)}>0$ and $[\text{Tr}(\Delta^\dagger\Delta)]^2>0$, we have
    \begin{equation}
    \label{con:vtr}
        \frac{V^{(4)}}{[\text{Tr}(\Delta^\dagger\Delta)]^2}>0~,
    \end{equation}
    so we can rewrite Eq.~\ref{con:frac} as follow,
    \begin{equation}
    \label{con:fr}
      f(r)\equiv ar^2+br+c>0~,
    \end{equation}
    where $a=\lambda_H, b=\lambda_1+\lambda_4\xi, c=\lambda_2+\lambda_3\zeta$. Obviously, Eq.~\ref{con:fr} is regarded as a one-dimensional quadratic inequality.
    
    (I) $b>0$
    
    For $f(r)>0$ and $r\in(0,\infty)$, the parabola opens up, $a>0$, which is
    \begin{equation}
        \lambda_H>0~.
    \end{equation}
    When $r\rightarrow0$, we have $c>0$ and
    \begin{equation}
    \label{con:f1}
        \lambda_2+\lambda_3\zeta>0~.
    \end{equation}
    In this case, Eq.~\ref{con:vtr} is apparently satisfied.
    
    (II)$b<0$
    
    Likewise,  $a>0$ and $c>0$ still hold. At this time, due to $f(r)>0$, the minimum must also satisfy this condition. For $f'(r)=0$, it can be obtained,
    \begin{equation}
        2\lambda_Hr+\lambda_1+\lambda_4\xi=0~,~ r=-\frac{\lambda_1+\lambda_4\xi}{2\lambda_H}~.
    \end{equation}
    Substitute the minimum point back to Eq.\ref{con:frac}, 
    \begin{equation}
        -(\lambda_1+\lambda_4\xi)^2+4\lambda_H(\lambda_2+\lambda_3\zeta)>0~.
    \end{equation}
    Simplified available
    \begin{equation}
        -2\sqrt{\lambda_H(\lambda_2+\lambda_3\zeta)}<\lambda_1+\lambda_4\xi<2\sqrt{\lambda_H(\lambda_2+\lambda_3\zeta)}~.
    \end{equation}
    Since the case where $b$ is less than zero is considered here, it is only necessary to consider that
    \begin{equation}
        -2\sqrt{\lambda_H(\lambda_2+\lambda_3\zeta)}<\lambda_1+\lambda_4\xi~,
    \end{equation}
    which is
    \begin{equation}
    \label{con:f2}
        \lambda_1+\lambda_4\xi+2\sqrt{\lambda_H(\lambda_2+\lambda_3\zeta)}>0~.
    \end{equation}
    To sum up, we can make a summary and redefine Eq.~\ref{con:f1} and Eq.~\ref{con:f2},
    \begin{align}
    \label{con:f3}
     &\lambda_H>0~,\\
     &\lambda_2+\lambda_3\zeta\equiv F_1(\zeta)>0~,\\
     &\lambda_1+\lambda_4\xi+2\sqrt{\lambda_H(\lambda_2+\lambda_3\zeta)}\equiv F_2(\xi,\zeta)>0~.
    \end{align}
    Since the value range of $\zeta$ is $2\xi^2-2\xi+1\leqslant\zeta\leqslant1$, which is proofed in Appendix.~\ref{con:proof}, and $F_2(\xi,\zeta)$ is monotonic about $\zeta$ and $\xi$, we can take $\zeta=2\xi^2-2\xi+1$ and rewrite Eq.~\ref{con:f3} as the function $F(\xi)$
    \begin{equation}
    \label{con:fxi}
        F(\xi)=\lambda_1+\xi\lambda_4+2\sqrt{\lambda_H[\lambda_2+\lambda_3(2\xi^2-2\xi+1)]}~.
    \end{equation}
    Since $\xi\in[0,1]$ and $F(\xi)>0$, there are $F(0)>0$ and $F(1)>0$, i.e.
    \begin{equation}
    \label{eq:f0}
        F(0)=\lambda_1+2\sqrt{\lambda_H(\lambda_2+\lambda_3)}>0~,
    \end{equation}
    \begin{equation}
    \label{eq:f1}
        F(1)=\lambda_1+\lambda_4+2\sqrt{\lambda_H(\lambda_2+\lambda_3)}>0~.
    \end{equation}
    If $F(\xi)$ is a monotonic function, the above two conditions are sufficient. On the contrary, it is necessary to find its minimum point. So we need to take the first derivative of $F(\xi)$,
    \begin{equation}
    \label{con:firstde}
        F'(\xi)=\lambda_4+\frac{2\lambda_H\lambda_3(2\xi-1)}{\sqrt{\lambda_H[\lambda_2+\lambda_3(2\xi^2-2\xi+1)]}}~,
    \end{equation}
    whose second derivative is
    \begin{equation}
    \label{con:secondde}
        F''(\xi)=\frac{2\lambda_H^2\lambda_3(2\lambda_2+\lambda_3)}{{\lambda_H[\lambda_2+\lambda_3(2\xi^2-2\xi+1)]}^{3/2}}~.
    \end{equation}
    It can be seen from the above formula that the sign of $F''(\xi)$ is the same as that of $\lambda_3$. If $\lambda_3>0$ and $F'(0)<0,~F'(1)>0$, the minimum falls in $(0,1)$. Otherwise, the minimum value is $F(0)$ or $F(1)$, and Eq.~\ref{eq:f0} with Eq.~\ref{eq:f1} is sufficient to ensure $F(\xi)>0$. For $F'(0)<0$ and $F'(1)>0$, we have
    \begin{align}
    \label{mini0}
        &F'(0)=\lambda_4-\frac{2\lambda_H\lambda_3}{\sqrt{\lambda_H(\lambda_2+\lambda_3)}}<0~,\\
    \label{mini1}
        &F'(1)=\lambda_4+\frac{2\lambda_H\lambda_3}{\sqrt{\lambda_H(\lambda_2+\lambda_3)}}>0~.
    \end{align}
    Combining Eq.~\ref{mini0} and Eq.~\ref{mini1} gives
    \begin{equation}
        2\lambda_3\sqrt{\lambda_H}>|\lambda_4|\sqrt{\lambda_2+\lambda_3}~.
    \end{equation}
    That is to say, if and only if $ 2\lambda_3\sqrt{\lambda_H}>|\lambda_4|\sqrt{\lambda_2+\lambda_3}$, we should calculate the minimum of $F(\xi)$ in $(0,1)$. Taking zero for Eq.~\ref{con:firstde} and solving the equation, the extreme point can be obtained as
    \begin{equation}
        \xi_0=\frac{1}{2}-\frac{\lambda_4}{2\lambda_3}\sqrt{\frac{\lambda_3(2\lambda_2+\lambda_3)}{8\lambda_H\lambda_3-\lambda_4^2}}~.
    \end{equation}
    Substituting $\xi_0$ back into Eq.~\ref{con:fxi}, the $F(\xi)>0$ condition can be obtained
    \begin{equation}
    \begin{aligned}
        F(\xi_0)&=\lambda_1+\frac{1}{2}\lambda_4+4\lambda_H\sqrt{\frac{\lambda_3(2\lambda_2+\lambda_3)}{8\lambda_H\lambda_3-\lambda_4^2}}-\frac{\lambda_4^2}{2\lambda_3}\sqrt{\frac{\lambda_3(2\lambda_2+\lambda_3)}{8\lambda_H\lambda_3-\lambda_4^2}}\\
        &=\lambda_1+\frac{1}{2}\lambda_4+\frac{1}{2}(8\lambda_H\lambda_3-\lambda_4^2)\sqrt{\frac{2\frac{\lambda_2}{\lambda_3}+1}{8\lambda_H\lambda_3-\lambda_4^2}}\\
        &=\lambda_1+\frac{1}{2}\lambda_4+\frac{1}{2}\sqrt{(8\lambda_H\lambda_3-\lambda_4^2)(2\frac{\lambda_2}{\lambda_3}+1)}>0~.
    \end{aligned} 
    \end{equation}
    
    In summary, the vacuum stability conditions are obtained as follow
    \begin{equation}
    \lambda_H,\lambda_2+\lambda_3,2\lambda_2+\lambda_3,\lambda_1+2\sqrt{\lambda_H(\lambda_2+\lambda_3)},\lambda_1+\lambda_4+2\sqrt{\lambda_H(\lambda_2+\lambda_3)}>0~,
    \end{equation}
    and
    \begin{equation}
        2\lambda_3\sqrt{\lambda_H}\leqslant\lvert\lambda_4\rvert\sqrt{\lambda_2+\lambda_3}~\text{or}~\lambda_1+\frac{1}{2}\lambda_4+\frac{1}{2}\sqrt{(8\lambda_H\lambda_3-\lambda_4^2)(2\frac{\lambda_2}{\lambda_3}+1)}>0~.
    \end{equation}
    
    \section{Proof of the properties of $\xi$ and $\zeta$}
    \label{con:proof}
    \subsection{$0\leq\xi\leq1$}
    We already know
    \begin{equation}
    \label{con:matrix}
        \Delta=\left(
    \begin{matrix}
        \Delta^+/\sqrt{2} & \Delta^{++}\\
        \Delta^0 & -\Delta^+/\sqrt{2}
    \end{matrix}
    \right),~
    \Delta^\dagger=\left(
    \begin{matrix}
        \Delta^-/\sqrt{2} & \Delta^0\\
        \Delta^{--} & -\Delta^-/\sqrt{2}
    \end{matrix}
    \right)~.
    \end{equation}
    From the above formula, we can easily find
    \begin{equation}
        \text{Tr}\Delta=0,~\text{Tr}\Delta^\dagger=0~.
    \end{equation}
    For any $2\times2$ matrices $M$ and $N$, we have
    \begin{equation}
        MN+NM=\boldsymbol1(\text{Tr}MN-\text{Tr}M\text{Tr}N)+M\text{Tr}N+N\text{Tr}M~,
    \end{equation}
    where $\boldsymbol{1}$ is identity matrix, from which follows immediately
    \begin{equation}
        \Delta\Delta^\dagger+\Delta^\dagger\Delta=\boldsymbol{1}\times\text{Tr}\Delta\Delta^\dagger~.
    \end{equation}
    Then multiply this by $H^\dagger$ on the left and right by $H$ to get
    \begin{equation}
        H^\dagger\Delta\Delta^\dagger H+H^\dagger\Delta^\dagger\Delta H=H^\dagger H\text{Tr}\Delta\Delta^\dagger~.
    \end{equation}
    Due to $H^\dagger\Delta^\dagger\Delta H\geq0$, we can obtain
    \begin{equation}
        H^\dagger H\text{Tr}\Delta\Delta^\dagger\geq H^\dagger\Delta\Delta^\dagger H~.
    \end{equation}
    According to the definition of $\xi$, it is easy to know
    \begin{equation}
        \xi\equiv\frac{H^\dagger\Delta\Delta^\dagger H}{H^\dagger H\text{Tr}\Delta\Delta^\dagger}\leq1~.
    \end{equation}
    Since $H^\dagger H\text{Tr}\Delta\Delta^\dagger$ and $H^\dagger\Delta\Delta^\dagger H$ are positive, $\xi$ must be trivially greater than zero. Finally the two boundary values 0 and 1 are effectively reached respectively when $H^\dagger\Delta=0$ and $\Delta H=0$. Thus
    \begin{equation}
        0\leq\xi\leq1~.
    \end{equation}
    \subsection{$\frac{1}{2}\leq\zeta\leq1$}
    Using Eq.~\ref{con:matrix} and considering their product
    \begin{equation}
    \label{matrix}
        \Delta\Delta^\dagger=\left(
    \begin{matrix}
        {\lvert\frac{\Delta^+}{\sqrt{2}}\rvert}^2+{\lvert\Delta^{++}\rvert}^2 & \frac{\Delta^+\Delta^0}{\sqrt{2}}+\frac{\Delta^{++}\Delta^-}{\sqrt{2}}\\
        \frac{\Delta^-\Delta^0}{\sqrt{2}}-\frac{\Delta^{--}\Delta^+}{\sqrt{2}} & {\lvert\frac{\Delta^+}{\sqrt{2}}\rvert}^2+{\lvert\Delta^0\rvert}^2
    \end{matrix}
    \right)~.
    \end{equation}
    First, trace Eq.~\ref{matrix} and then square it, we have
    \begin{equation}
    \label{trace1}
         (\text{Tr}\Delta\Delta^\dagger)^2=({\lvert\frac{\Delta^+}{\sqrt{2}}\rvert}^2+{\lvert\Delta^{++}\rvert}^2)^2+({\lvert\frac{\Delta^+}{\sqrt{2}}\rvert}^2+{\lvert\Delta^{++}\rvert}^2)^2+2({\lvert\frac{\Delta^+}{\sqrt{2}}\rvert}^2+{\lvert\Delta^{++}\rvert}^2)({\lvert\frac{\Delta^+}{\sqrt{2}}\rvert}^2+{\lvert\Delta^0\rvert}^2)~.
    \end{equation}
    Furthermore, square Eq.~\ref{matrix} and then find the trace to get
    \begin{equation}
    \label{trace2}
        \text{Tr}(\Delta\Delta^\dagger)^2=({\lvert\frac{\Delta^+}{\sqrt{2}}\rvert}^2+{\lvert\Delta^{++}\rvert}^2)^2+({\lvert\frac{\Delta^+}{\sqrt{2}}\rvert}^2+{\lvert\Delta^{++}\rvert}^2)^2+2[\frac{\Delta^+\Delta^0}{\sqrt{2}}+\frac{\Delta^{++}\Delta^-}{\sqrt{2}}][\frac{\Delta^-\Delta^0}{\sqrt{2}}-\frac{\Delta^{--}\Delta^+}{\sqrt{2}}]~.
    \end{equation}
    Finally, subtract Eq.~\ref{trace1} and Eq.~\ref{trace2},
    \begin{equation}
    \label{trace3}
    \begin{aligned}
         (\text{Tr}\Delta\Delta^\dagger)^2-\text{Tr}(\Delta\Delta^\dagger)^2&=2({\lvert\frac{\Delta^+}{\sqrt{2}}\rvert}^2+{\lvert\Delta^{++}\rvert}^2)({\lvert\frac{\Delta^+}{\sqrt{2}}\rvert}^2+{\lvert\Delta^0\rvert}^2)\\
         &-2[\frac{\Delta^+\Delta^0}{\sqrt{2}}+\frac{\Delta^{++}\Delta^-}{\sqrt{2}}][\frac{\Delta^-\Delta^0}{\sqrt{2}}-\frac{\Delta^{--}\Delta^+}{\sqrt{2}}]\\
         &=2\text{Det}\Delta\Delta^\dagger
    \end{aligned}~.
    \end{equation}
    Using $\text{Det}\Delta\Delta^\dagger\equiv|\text{Det}\Delta|^2\geq0$ and Eq.~\ref{trace3}, it is easily to note that $(\text{Tr}\Delta\Delta^\dagger)^2\geq\text{Tr}(\Delta\Delta^\dagger)^2$. According to the definition of $\zeta$, we have
    \begin{equation}
        \zeta\equiv\frac{\text{Tr}(\Delta\Delta^\dagger)^2}{(\text{Tr}\Delta\Delta^\dagger)^2}\leq1~,
    \end{equation}
    and $\zeta$ must be positive because $(\text{Tr}\Delta\Delta^\dagger)^2$ and $\text{Tr}(\Delta\Delta^\dagger)^2$ are both greater than zero. But in fact, $\zeta$ cannot go below 1/2. To see this, we will rewrite $\zeta$ by $M_1^2$ and $M_2^2$, the two (real and positive) eigenvalues of $\Delta\Delta^\dagger$,
    \begin{equation}
        \zeta=\frac{M_1^4+M_2^4}{(M_1^2+M_2^2)^2}~,
    \end{equation}
    and then divide it up and down by $M_2^4$, we can easily read the function $\zeta(x)=(1+x^2)/(1+x)^2$ where $x\equiv M_1^2/M_2^2\geq0$ to show that it has a minimum of $\zeta=1/2$ when $x=1$. Similarly, we calculated that $\zeta\leq1$ and reaches 1 at $x\rightarrow0$ or $x\rightarrow\infty$. Therefore,
    \begin{equation}
        \frac{1}{2}\leq\zeta\leq1~.
    \end{equation}
    \subsection{Correlation between $\xi$ and $\zeta$}
    We first perform a general gauge transformation $H\rightarrow\mathcal{U}H$, $\Delta\rightarrow\mathcal{U}\Delta\mathcal{U}^\dagger$, where $\mathcal{U}$ is any element of $SU(2)_L\times U(1)_Y$. Since $\mathcal{U}$ is unitary and $\Delta\Delta^\dagger$ is hermitian, we can always find a gauge transformation that diagonalizes $\Delta\Delta^\dagger$ for any given field $\Delta$. According to the definition of $\xi$ and a series of calculation, $\xi$ reads
    \begin{equation}
        \xi=\frac{M_2^2|\Tilde{\phi^0}|^2+M_1^2|\Tilde{\phi^+}|^2}{(M_1^2+M_2^2)(|\Tilde{\phi^0}|^2+|\Tilde{\phi^+}|^2)}~,
    \end{equation}
    where ${\rm diag}\{ M_1, M_2 \}$ is the diagonalized matrix of $\Delta$ and the tilde denotes the components of the transformed doublet $\Tilde{H}=\mathcal{U}H$. Furthermore, we define
    \begin{align}
        &c_\Delta^2\equiv\frac{M_1^2}{M_1^2+M_2^2},~s_\Delta^2\equiv\frac{M_2^2}{M_1^2+M_2^2}~,\\
        &c_H^2\equiv\frac{|\Tilde{\phi^+}|^2}{|\Tilde{\phi^0}|^2+|\Tilde{\phi^+}|^2},~s_H^2\equiv\frac{|\Tilde{\phi^0}|^2}{|\Tilde{\phi^0}|^2+|\Tilde{\phi^+}|^2}~,
    \end{align}
    they scan all their allowed domains $c_\Delta^2(\text{or} s_\Delta^2), c_H^2(\text{or} s_H^2)\in[0,1]$. Using these definitions, we can rewrite $\xi$ and $\zeta$ as
    \begin{equation}
    \label{rewritexi}
    \begin{aligned}
        \xi=&\frac{M_1^2\lvert\Tilde{\phi^+}\rvert^2+M_2^2\lvert\Tilde{\phi^0}\rvert^2}{(M_1^2+M_2^2)(\lvert\Tilde{\phi^+}\rvert^2+\lvert\Tilde{\phi^0}\rvert^2)}\\
        =&c_\Delta^2c_H^2+s_\Delta^2s_H^2\\
        =&\frac{1}{2}[(c_\Delta^2+s_\Delta^2)(c_H^2+s_H^2)+(c_\Delta^2-s_\Delta^2)(c_H^2-s_H^2)]\\
        =&\frac{1}{2}(1+c_{2\Delta}c_{2H})~,
    \end{aligned}
    \end{equation}
    \begin{equation}
    \label{rewritezeta}
    \begin{aligned}
        \zeta=&\frac{M_1^4+M_2^4}{(M_1^2+M_2^2)^2}=(\frac{M_1^2}{M_1^2+M_2^2})^2+(\frac{M_2^2}{M_1^2+M_2^2})^2\\
        =&c_\Delta^4+s_\Delta^4=\frac{1}{2}[(c_\Delta^2+s_\Delta^2)^2+(c_\Delta^2-s_\Delta^2)^2]\\
        =&\frac{1}{2}(1+c_{2\Delta}^2)~,
    \end{aligned}    
    \end{equation}
    where we have defined $c_{2H}=c_H^2-s_H^2$, $c_{2\Delta}=c_\Delta^2-s_\Delta^2$ with their range of variation $c_{2H}\in[-1,1]$, $c_{2\Delta}\in[-1,1]$. Combining Eq.~\ref{rewritexi} and Eq.~\ref{rewritezeta}, and then eliminating $c_{2\Delta}^2$, we can obtain
    \begin{equation}
        2\xi^2-2\xi+\frac{c_{2H}^2+1}{2}=c_{2H}^2\zeta~.
    \end{equation}
    In other words, we have
    \begin{equation}
        2\xi^2-2\xi+\frac{1}{2}=(\zeta-\frac{1}{2})c_{2H}^2~.
    \end{equation}
    Since the value of $c_{2H}^2\in[0,1]$, it is easy to get
    \begin{equation}
        \zeta\geq2\xi^2-2\xi+1~.
    \end{equation}
    It is obvious to find that $2\xi^2-2\xi+1\geq\frac{1}{2}$ with $\xi\in[0,1]$, and as mentioned earlier, $\zeta\in[\frac{1}{2},1]$, so we can obtain
    \begin{equation}
        2\xi^2-2\xi+1\leqslant\zeta\leqslant1~.
    \end{equation}
    
    \section{1-loop renormalization group equations}
    \label{rge_oneloop}
    The 1-loop renormalization group equations are extracted below. For simplicity, among the SM Yukawa couplings, we only retain the top coupling. Similarly, we take the normal ordering assumption of neutrino mass spectrum and neglect the contribution of the first 2 generation neutrino Yukawa coupling. Here $t=\log \mu$.
    \begin{align*}
    \label{con:rges}
    (4\pi)^2\frac{dg_1}{dt}=&\frac{47}{10}g_1^3~,
    \\   
    (4\pi)^2\frac{dg_2}{dt}=&-\frac{5}{2}g_2^3~,
    \\   
    (4\pi)^2\frac{dg_3}{dt}=&-7g_3^3~,
    \\
    (4\pi)^2\frac{dy_t}{dt}=&-\frac{17}{20}g_1^2y_t-\frac{9}{4}g_2^2y_t-8g_3^2y_t+\frac{9}{2}y_t^3~,
    \\
    (4\pi)^2\frac{dy_\nu}{dt}=&-\frac{9}{10}g_1^2y_\nu-\frac{9}{2}g_2^2y_\nu+2y_\nu^3~,
    \\
    (4\pi)^2\frac{d\lambda_H}{dt}=&\frac{27}{200}g_1^4+\frac{9}{20}g_1^2g_2^2+\frac{9}{8}g_2^4+3\lambda_1^2+3\lambda_1\lambda_4+\frac{5}{4}\lambda_4^2-\frac{9}{5}g_1^2\lambda_H-9g_2^2\lambda_H\\
    &+24\lambda_H^2+12\lambda_Hy_t^2-6y_t^4~,
    \\
    (4\pi)^2\frac{d\lambda_1}{dt}=&\frac{27}{25}g_1^4-\frac{18}{5}g_1^2g_2^2+6g_2^4-\frac{9}{2}g_1^2\lambda_1-\frac{33}{2}g_2^2\lambda_1+4\lambda_1^2+16\lambda_1\lambda_2+12\lambda_1\lambda_3\\
    &+6\lambda_2\lambda_4+2\lambda_3\lambda_4+\lambda_4^2+12\lambda_1\lambda_H+4\lambda_H\lambda_4+6\lambda_1y_t^2+\lambda_1y_\nu^2~,
    \\   
    (4\pi)^2\frac{d\lambda_2}{dt}=&\frac{54}{25}g_1^4+15g_2^4+2\lambda_1^2-24g_2^2\lambda_2+28\lambda_2^2-\frac{36}{5}g_1^2(g_2^2+\lambda_2)+24\lambda_2\lambda_3\\
    &+6\lambda_3^2+2\lambda_1\lambda_4+2\lambda_2y_\nu^2~,
    \\
    (4\pi)^2\frac{d\lambda_3}{dt}=&\frac{72}{5}g_1^2g_2^2-6g_2^4-\frac{36}{5}g_1^2\lambda_3-24g_2^2\lambda_3+24\lambda_2\lambda_3+18\lambda_3^2+\lambda_4^2+2\lambda_3y_v^2-y_\nu^4~,
    \\
    (4\pi)^2\frac{d\lambda_4}{dt}=&\frac{36}{5}g_1^2g_2^2-\frac{9}{2}g_1^2\lambda_4-\frac{33}{2}g_2^2\lambda_4+8\lambda_1\lambda_4+4\lambda_2\lambda_4+8\lambda_3\lambda_4+4\lambda_4^2+4\lambda_H\lambda_4\\
    &+6\lambda_4y_t^2+\lambda_4y_\nu^2~.
    \end{align*}

\end{document}